\documentstyle[12pt]{article}

\newcommand{\EQ}{\begin{equation}}
\newcommand{\EN}{\end{equation}}
\def\aprle{\buildrel < \over {_{\sim}}}
\def\aprge{\buildrel > \over {_{\sim}}}
\begin{document}
\topmargin 0pt
\oddsidemargin=-0.4truecm
\evensidemargin=-0.4truecm      
\renewcommand{\thefootnote}{\fnsymbol{footnote}}     
\newpage
\setcounter{page}{0}
\begin{titlepage}     
\begin{flushright}
SISSA 40/96/A-EP \\
TUM-HEP--239/96\\
SFB-375/83
\end{flushright}
\begin{center}
{\large RESONANT NEUTRINO SPIN--FLAVOR PRECESSION\\
AND SUPERNOVA SHOCK REVIVAL}\\
\vspace{0.5cm}
{\large E.Kh. Akhmedov  
\footnote{On leave from NRC ``Kurchatov Institute'', Moscow 123182, Russia}}\\ 
{\em
Institut f\"ur Theoretische Physik,
Technische Universit\"at M\"unchen,          \\
James--Franck--Strasse, D--85748 Garching, Germany}
\vskip .5cm
{\large A. Lanza,  ~S.T. Petcov, \footnote{Istituto Nazionale di Fisica 
Nucleare, Sezione di Trieste, Italy}
\footnote{Institute of Nuclear Research and Nuclear Energy, Bulgarian 
Academy of Sciences, BG-1784 
Sofia, Bulgaria}, and D.W. Sciama\footnote{International Centre for 
Theoretical Physics, Strada Costiera 11, I-34100 Trieste, Italy}}\\
{\em Scuola Internazionale Superiore di Studi Avanzati\\
Via Beirut 2--4, I-34014 Trieste, Italy} \\
\end{center}
\begin{abstract}
A new mechanism of supernova shock revival is proposed, which 
involves resonant spin--flavor precession of neutrinos with a transition 
magnetic moment in the magnetic field of the supernova.
The mechanism can be operative in supernovae for transition magnetic 
moments as small as $10^{-14}\mu_B$ provided the neutrino 
mass squared difference is in the range $\Delta m^2 \sim  
(3 \;{\rm eV})^2-(600 \;{\rm eV})^2$. It is shown that this mechanism 
can increase the neutrino--induced shock reheating energy by about 60\%.  

\end{abstract}
\end{titlepage}
\renewcommand{\thefootnote}{\arabic{footnote}}
\setcounter{footnote}{0}
\newpage
\section{Introduction}  

One of the most important problems in the theory of supernova explosions 
is to understand the physical mechanisms which eventually expel the outer
mantle delivering the right amount of energy. Although the main ideas
involved in the theory have received major confirmation after the detection
of the neutrinos from SN1987A, the problem of accelerating the outward
going shock wave
which forms after core collapse is still unresolved. For many years,
when all computer calculations were essentially done in one dimension, the
majority of computations were unsuccessful since the shock would travel
for about $300-500$ km and then stall, after losing 
energy by dissociating heavy nuclei in the envelope into nucleons. 
Many proposals have been made to solve this problem, starting from
including general relativistic corrections, different equations of state,
better neutrino transport description and new neutrino physics. 
Recent numerical calculations \cite{Her94,Bur95,Janka} 
have shown that as one moves to more than one dimension one
can easily get convective instabilities
driven by neutrino heating which are very effective and fast in reheating
the material behind the shock. The idea that convection would help the
explosion is not new. It has been in the literature for some time,
see, e.g.,  \cite{Epst79} (see \cite{Bur95} 
for an extensive list of references). The idea that multidimensional
calculations might help is also not new. Several attempts have been made 
to model supernova explosion in more than one dimension; probably they 
failed because they did not contain the right combinations of other 
aspects of the physics involved. Although these most recent calculations 
make an important contribution to the subject we still 
believe that much work should be done to understand fully the mechanisms
involved. The proposed convective instability relies heavily on the 
details of neutrino interactions with matter which control the 
energy transport 
process. The details of this process are still controversial, especially 
when the matter is at high density and is not spherically symmetrical. 
The question whether convective instabilities can revive the 
shock and lead to a successful supernova explosion is therefore 
still far from settled, and any new mechanism which could contribute 
towards the shock energy would be very welcome.  

Recently there has been discussion of whether massive 
neutrinos, which seem to be necessary to reconcile the solar neutrino 
experiments with the standard model of the sun, might also help
in accelerating the shock. Matter--enhanced neutrino oscillations 
(MSW effect, \cite{Wolf78,Wolf79,MS85}) in supernovae might play an 
important role in reviving the shock after core collapse by increasing the 
amount of energy that neutrinos deposit behind the shock 
\cite{MS86,Ful87,Ful92}. The idea is based on the fact that in the region 
between the neutrinosphere and the position of the stalled
shock the matter density is such that flavor transformation of $\nu_\mu$ 
or $\nu_\tau$ into $\nu_e$ is resonant for masses of the heavier 
neutrinos in 
the range $10-100$ eV; this transformation can be efficient even if the 
vacuum neutrino mixing angle is quite small, $\theta\aprge 10^{-4}$. 
Since the average energy of $\nu_\mu$'s and $\nu_\tau$'s at the 
neutrinosphere is about $20$ MeV whereas that of $\nu_e$'s is about $10$ 
MeV, the electron neutrinos emerging as a result of the $\nu_{\mu}
(\nu_{\tau})\to \nu_e$ transformation would have twice as high an energy 
as the originally emitted ones, and this extra energy would be available for 
heating the matter behind the shock. Electron neutrinos interact with matter 
with a larger 
cross section than muon or tauon neutrinos since they have charged--current 
interactions with matter in addition to the neutral--current ones. 
Therefore the $\nu_e$'s produced in the $\nu_{\mu}(\nu_{\tau})\to 
\nu_e$ transformation will more efficiently deposit energy behind the shock. 
Fuller et al. \cite{Ful92} have shown 
that the net effect is a $\sim 60\%$ increase in the supernova explosion 
energy.

In this paper, we show that a similar result can be obtained in the framework
of the resonant spin--flavor precession mechanism. In this case one has
to assume that the neutrino has a nonzero transition magnetic moment $\mu$
by which it interacts with a magnetic field. Since we know that 
after a supernova explosion a pulsar is, in many cases, left over, the 
important role played by the magnetic field during and after core
collapse is not in question. The strong magnetic field and the high
density make the environment between the neutrinosphere and the position
of the stalled shock suitable for spin--flavor conversion due to 
transition magnetic moments of neutrinos. 

Very much as in the case of the MSW effect, spin--flavor precession due to 
a transition magnetic moment of neutrinos, in which neutrino helicity and 
flavor are rotated simultaneously \cite{ShVa}, can be resonantly enhanced in 
matter \cite{Akhm88,LM88}. This effect can explain the observed deficit 
of solar neutrinos with respect to the predictions of the standard solar 
model \cite{Akhm88,LM88,ALP93,LNPul,KraOT,ALP95}. That requires the neutrino 
transition magnetic moment to be of the order of $\mu\approx 10^{-11}\mu_B$ 
($\mu_B=e/2m_e$ is the electron Bohr magneton) provided the strength of 
the magnetic field near the bottom of the convective zone of the sun is 
of the order of a few tens of kG. The indicated value of the transition 
magnetic moment is to be compared with recent astrophysical upper bounds 
derived from the limits on the energy loss rates of white dwarfs 
($10^{-11}\mu_B$, ref. \cite{Blin}) and helium stars ($3\times 
10^{-12}\mu_B$, ref. \cite{Raff90}, and $10^{-12}\mu_B$, ref. 
\cite{CastDInn}). The latter two values imply that an order of magnitude 
stronger magnetic field might be necessary to account for the solar neutrino 
problem in the framework of the neutrino magnetic moment scenario.  

In the present paper we show that the spin--flavor precession of neutrinos may 
play an important role in supernova dynamics even if neutrino transition 
magnetic moments are far below the present astrophysical upper limits.  
In particular, it can be 
resonantly enhanced in the region between the neutrinosphere and the  
position of the shock for typical values of $\mu\approx 10^{-14}\mu_B$, 
magnetic field strengths of $B\approx 10^{12}-10^{15}$ G and neutrino 
masses which lie in the range $\sim (3-600)$ eV. This strength of the 
magnetic field is natural in the context of supernovae if the explosion 
does give rise to a pulsar, the value of $\mu$ is consistent
with the prediction of the decaying neutrino hypothesis
\cite{Sciama1993book}, and the range of neutrino masses is
the one which is relevant for cosmology and the decaying
neutrino theory. 

The idea that neutrino magnetic moments can play 
an important role in supernova dynamics was first put forward by 
Dar \cite{Dar} in the context of the usual Dirac neutrino magnetic moments 
and transitions of active left-handed neutrinos into sterile right-handed 
ones. Our mechanism, based on the spin--flavor precession of neutrinos due 
to their Majorana-like transition magnetic moments, is different from 
Dar's. Resonant spin--flavor precession (RSFP) of neutrinos in type II 
supernovae has been studied earlier \cite{AB,APS}; however the main goal 
of those papers was to explore possible consequences for the neutrino signal 
from supernovae, and no implications of this effect for supernova dynamics 
were discussed. 
 
The plan of the paper is as follows. We start by reviewing the main 
features of RSFP in Sec. 2. In Sec. 3 we discuss the RSFP in supernovae, 
and in Sec. 4 consider the implications of this effect for supernova shock 
reheating. Sec. 5 contains our conclusions. 

\section{Basic features of RSFP}

We confine our analysis to the simplest case 
of Majorana neutrinos, for which the diagonal magnetic moments 
are zero. We also assume that there are only two neutrino flavors and 
disregard the usual flavor mixing, i.e. we consider transitions in a 
two-neutrino system  $\nu_e$ and $\nu_a$ ($\nu_a$ can be either 
$\nu_\tau$ or $\nu_\mu$) with masses $m_{\nu_e}$ and $m_{\nu_a}$ due to the 
transition magnetic moment $\mu_{ea}\equiv \mu$.
More specifically, of main 
interest to us are the transitions between the 
right--handed electron antineutrinos $\bar\nu_{eR}$ and 
left--handed muon or tauon neutrinos $\nu_{\mu L}$ or $\nu_{\tau L}$ 
($\nu_{aL}$). As we shall see, the transitions between the corresponding 
antiparticles, namely $\nu_{eL} \leftrightarrow \bar{\nu}_{\mu R}
(\bar{\nu}_{\tau R})$, which may be relevant for the solution of the solar 
neutrino problem, are non-resonant in the supernova environment behind 
the shock provided $m_{\nu_a} > m_{\nu_e}$. We shall comment on the case 
$m_{\nu_a} < m_{\nu_e}$ in Sec. 4.  

In a medium mainly composed of electrons, neutrons and protons the evolution 
of the $\bar{\nu}_{eR}$ and $\nu_{aL}$ states in a transverse magnetic 
field $B_\bot$ is described by the Schr\"{o}dinger--like equation  
\cite{Akhm88,LM88}
\EQ
i\frac{d}{dt}\left(\begin{array}{c}\bar{\nu}_{eR}\\ {\nu}_{aL }\end{array}
\right)=\left(\begin{array}{cc}
-A  & B\\
 B & A
\end{array}\right)
\left(\begin{array}{c}\bar{\nu}_{eR}\\ {\nu}_{aL}\end{array}\right)
\label{evol}
\EN
where $A=\Delta m^2/4E +V/2$, $B=\mu B_\bot$, $\Delta m^2
\equiv m_{\nu_a}^2-m_{\nu_e}^2$ and $V$ is the difference of the effective 
potentials 
that neutrinos $\nu_{aL}$ and $\bar{\nu}_{eR}$ experience in matter. The 
effective 
potential $V$ can be written as a sum of two contributions, $V=V^{(1)}+
V^{(2)}$. Numerically the most important term $V^{(1)}$ is due to the 
interaction of 
neutrinos with matter: $V^{(1)}=\sqrt{2}G_F(N_e-N_n)$ where $G_F$ is 
the Fermi constant and $N_e$ and $N_n$ are the electron and
neutron number densities respectively. The second term is 
due to neutrino--neutrino forward scattering, which in the case
of scattering of an electron neutrino by the neutrino sea
of all types takes the form \cite{Okun,BlinOkun,NoetzRaf} 
\EQ
V^{(2)}(\nu_e)=\sqrt{2}G_F(2 N_{\nu_e}^{eff}+N_{\nu_\mu}^{eff}
+N_{\nu_\tau}^{eff}), \label{V2}
\EN
where $N_{\nu_e}^{eff}$ is the difference between the effective number 
densities of electron neutrinos and antineutrinos, and similarly for 
$\nu_\mu$ and $\nu_\tau$. The word ``effective'' is used here because the 
neutrino densities are modified by the averaged factor 
$\langle 1-v_m\cos\alpha\rangle$ which takes into account the dependence of 
the corresponding contributions to the effective potential on the angle 
$\alpha$ between the momenta of interacting neutrinos. This effect is 
negligible for neutrino scattering on non-relativistic or randomly-moving 
particles, but leads to a strong suppression of the $V^{(2)}$ potential for 
neutrino scattering on relativistic neutrinos ($v_m=1$) moving nearly in the 
same direction \cite{NoetzRaf,Ful92,Qian93}. Expressions similar to eq. 
(\ref{V2}) hold also for the effective potentials of $\mu$ and  $\tau$ 
neutrinos. Antineutrinos will experience the opposite sign potential,  
$V^{(2)}(\bar\nu_l)=-V^{(2)}(\nu_l),~l=e,\mu,\tau$. In the case of 
$\nu_{\mu(\tau)L} \leftrightarrow \bar{\nu}_{eR}$ transitions, the effective 
potential $V$ reduces to 
\EQ
V=\sqrt{2} G_F [N_e-N_n+3N_{\nu_e}^{eff}+5N_{\nu_\mu}^{eff}], \label{V} 
\EN
where we have taken into account that in the supernova environment 
$N_{\nu_\mu}^{eff}=N_{\nu_\tau}^{eff}$. 
However, as we shall see, the neutrino-neutrino forward scattering 
contributions to $V$ are very small in the shock reheating epoch 
($t \simeq 0.15$ s after the shock bounce) and can be safely 
neglected. Typical average neutrino energies at this epoch are 
\cite{Ful92} $\langle E_{\nu_e}\rangle \approx 9$ 
MeV, $\langle E_{\bar\nu_e}\rangle \approx 12$ MeV and $\langle E_{\nu_\mu}
\rangle \approx\langle E_{\nu_\tau}\rangle \approx 20$ MeV.  The 
neutrinosphere is at $R_\nu\approx 50$ km and the shock position is at 
about $400$ km. The neutrino luminosities
at the shock reheating epoch are $L_{\nu_e}\approx L_{\bar\nu_e}
\approx L_{\nu_{\tau(\mu)}} \approx L_{\bar\nu_{\tau(\mu)}}\approx
5\times 10^{52}$ erg s$^{-1}$. By inserting these values into the 
expression for the effective number density (see eq. (5) in \cite{Qian93}) 
one gets 
\EQ
N_{\nu_e}^{eff}= 1.44\times 10^{34} \;{\rm cm}^{-3} 
\left(\frac {10 \; {\rm km}}{r} \right)^4, \;\;\; 
N_{\nu_\mu}^{eff}=N_{\nu_\tau}^{eff}=0, \label{N}
\EN
and therefore the effective potential due to neutrino--neutrino
scattering in the case of $\nu_{\mu(\tau)L} \leftrightarrow \bar{\nu}_{eR}$ 
transitions is
\EQ
V^{(2)}\approx -5.45\times 10^{-3} \left(\frac {10 {\rm \; km}}{r}
\right)^4\;{\rm eV}. \label{V2num}
\EN
This term is numerically very small compared to the main $V^{(1)}$ 
term in the region of interest to us, namely 50 km $\aprle r \aprle $ 400 
km. 

{}From eq. (\ref{evol}) one can find the resonance condition by equating  
the diagonal terms of the effective Hamiltonian of the neutrino system:
\EQ 
\sqrt{2}G_F\frac{\rho}{m_N} (1-2 Y_e)= \frac {\Delta m^2}{2E}, 
\label{res1} 
\EN
or
\EQ
7.54\times 10^{-14} \;{\rm eV}\;\rho (\frac{\rm g}{\rm cc})\,(1-2 Y_e)=
\frac {\Delta m^2}{2E}, \label{res2}
\EN
where $\rho$ is the matter mass density, $m_N$ is the mass of the 
nucleon, and $Y_e$ is the number of electrons per baryon.
For a given $\Delta m^2$ and energy $E$ eq. (\ref{res1}) gives the values
of the density at which the transition is resonant. Since in the 
region between the neutrinosphere and the shock position $Y_e$ is always 
less than 1/2 at the shock reheating epoch \cite{Ful92,Qian93}, from eqs. 
(\ref{res1}) or (\ref{res2}) it follows that for $\Delta m^2>0$ only the 
transitions $\nu_{\mu(\tau)L} \leftrightarrow \bar{\nu}_{eR}$ 
will be resonant, whereas the transitions between the corresponding 
antiparticles, for which the signs of the l.h.s. of eqs. (\ref{res1}) and 
(\ref{res2}) must be reversed, are non-resonant. 

The neutrino eigenstates in matter and a magnetic field are linear 
combinations of $\bar{\nu}_{eR}$ and 
${\nu}_{\mu (\tau)L}$ with the mixing angle defined through 
\EQ
\tan 2\theta = \frac{2\mu B_{\bot}}{-\sqrt{2}G_{F}(N_{n}-
N_{e})+\frac{\Delta m^2}{2E}}. \label{theta}
\EN
The efficiency of the $\nu_{\mu(\tau)L}\rightarrow \bar{\nu}_{eR}$ 
transition is determined by the degree of the adiabaticity which depends 
on both the neutrino energy and the magnetic field strength at the resonance:
\EQ
\gamma\equiv \pi \frac{\Delta r}{l_{r}}=
8\frac{E}{\Delta m^2}(\mu B_{\bot r})^{2}(L_{\rho})_r. \label{adiab}
\EN
Here $\Delta r$ is the resonance width, $l_{r}=\pi/\mu B_{\bot r}$ is
the precession length at the resonance and $L_{\rho}\equiv \vert
\frac {1}{\rho(1-2Y_e)} \frac {d\rho(1-2Y_e)}{dr}\vert^{-1}$ 
is the characteristic length over which the effective matter density 
$(1-2Y_e)\rho$ varies significantly in the supernova, $(L_{\rho})_r$ 
being its value at the resonance. 
For the RSFP to be efficient, $\gamma$ should be $\aprge 1$.

The probability of the $\nu_{\mu(\tau)L}\leftrightarrow \bar{\nu}_{eR}$ 
transition can be written in the following general form  \cite{Petcov}:  
\begin{eqnarray}
P(\nu_{\mu(\tau)} \rightarrow \bar\nu_e)= \frac {1}{2} -\frac{1}{2}\cos
2\theta_i\cos 2\theta_f(1-2P') - \sqrt{P'(1-P')}\cos 2\theta_i\sin 2\theta_f 
\nonumber \\
\times \cos (\Phi_{12} + \Phi_{22}) + 
 \sqrt{P'(1-P')}\sin 2\theta_i\cos 2\theta_f~\cos (\Phi_{12} - \Phi_{22}) 
\nonumber \\
 +\frac {1}{2} P'\sin 2\theta_i\sin 2\theta_f~(\cos 2\Phi_{12} 
 + \cos 2 \Phi_{22}) - \frac {1}{2} \sin 2\theta_i\sin 2\theta_f~
  \cos 2\Phi_{22}. \label{trans1}
\end{eqnarray}
Here $\theta_i$ and $\theta_f$ are the values of the 
neutrino mixing angle in matter at the initial and final points of the 
neutrino path in the magnetic field, which we will assume to be located at 
the surface of the neutrinosphere and far beyond the resonance at 
a density much smaller than the resonance density,  
$\Phi_{12}$ and $\Phi_{22}$ are two phases which are responsible for 
the possible oscillatory dependence of the probability (\ref{trans1}) on 
$E/\Delta m^2$, the magnetic field strength and matter density profiles, 
and $P'$ is the so-called ``jump'' probability , i.e., the probability of 
transition of one of the matter eigenstate neutrinos into another in the 
course of neutrino propagation in the magnetic field of the supernova. 
The value of the jump probability is therefore a measure of violation of the 
adiabaticity of neutrino propagation. As we shall argue in the next Section, 
for the values of $E/\Delta m^2$ and the magnetic field strengths of interest 
to us, $\theta_i\approx \pi/2$ and $\theta_f \approx 0$. Under these 
conditions the general expression for the transition probability 
(\ref{trans1}) simplifies considerably and reduces to 
\EQ
P(\nu_{\mu(\tau)} \rightarrow \bar\nu_e) \cong \frac {1}{2} -\frac{1}{2}
\cos 2\theta_i\cos 2\theta_f(1-2P'). \label{trans2}
\EN
For $P' \cong 0$ (adiabtic transitions) 
the transition probability can be very close to 1. 
For the purposes of our further analysis  the jump probability can be 
approximated by the Landau-Zener probability:
\EQ
P'\approx P_{LZ}=\exp(-\frac{\pi}{2}\gamma) . \label{LZ}
\EN
\section {RSFP in supernovae}

We know very little about the magnetic field in the
supernova environment. It is certain that at least
in those cases in which the explosion leaves a pulsar, the
magnetic fields may be as strong as $10^{12} - 10^{14}$ G. 
However, also in these cases nothing is known about the
spatial distribution of the field. We shall assume rather 
arbitrarily that the radial dependence of the magnetic field strength 
above the neutrinosphere has the power-law behavior  
\EQ
B_\bot (r)=B_0 \left (\frac {r_0}{r}\right )^k, \;\; r\ge r_0\;,
\label{B}
\EN
where $r_0$ is the radius of the neutrinosphere, $B_0$ is
the strength of the field at the neutrinosphere, and $k=$2 or 3.

For a given $\Delta m^2/E$, the resonance condition (\ref{res2}) 
determines the value of the matter density (and, given the density profile, 
of the radial coordinate) at which the transition is most efficient.  
In order for the resonance to occur between the neutrinosphere  
which is located at about $50$ km and the position of the stalled shock 
$\sim 400$ km, $\Delta m^2/E$ should be in the range $5.5\times 10^{-7}$ 
to $2\times 10^{-2}$ eV. For typical neutrino energies $\langle E_{\nu_\mu}
\rangle \approx\langle E_{\nu_\tau}\rangle \approx 20$ MeV this 
would correspond to values of $\Delta m^2$ in the range 11 eV$^2$ to 
$4\times 10^5$ eV$^2$. If we assume the hierarchical pattern of neutrino 
masses, this would mean that the mass of the heavier neutrino should lie 
in the range $\sim$ 3 eV to 600 eV. If the shock stalls at larger distances 
from the supernova core, even smaller neutrino masses would do. 

For the above values of $\Delta m^2/E$ one 
can readily calculate the magnitudes of the product $\mu B_\bot$ which are 
necessary in order to have an adiabatic or weakly nonadiabatic transition 
(i.e. $\gamma \ge 1$). The values of $L_\rho$ at different resonance 
positions can be read off from Fig. 1. One then gets from eq. (\ref{adiab}) 
\EQ 
\mu B_{\bot r} \ge 3.5\times 10^{-7} \; {\rm eV} \label{lim1}
\EN
if the resonance takes place at $50$ km, or
\EQ
\mu B_{\bot r} \ge 1.1\times 10^{-9} \; {\rm eV} \label{lim2} 
\EN
when the resonance takes place at $350$ km. 
Assuming the transition magnetic moment $\mu=10^{-14}\;(10^{-12})\mu_B$, this 
gives 
the following lower bounds on the magnetic field strength at the resonance: 
\begin{eqnarray}
B_{\bot r} \ge 6.0\times 10^{15}\;(6.0\times 10^{13})\; {\rm G}\;\; 
& (r_{res}\simeq 50\; {\rm km});\nonumber \\
B_{\bot r} \ge 1.9\times 10^{13}\;(1.9\times 10^{11})\; {\rm G}\;\; 
& (r_{res}\simeq 350\; {\rm km}). \label{limB}
\end{eqnarray}
Alternatively, if one assumes the magnetic field strength at the 
neutrinosphere $B_0=5\times 10^{14}$ G and $k=2$, eqs. (\ref{lim1}) and 
(\ref{lim2})  transform into the following lower limit for the transition 
magnetic moment $\mu$: 
\EQ
\mu \ge 10^{-14}\;\; {\rm to}\;\; 10^{-13} \mu_B, \label{limmu}
\EN
which is one to two orders of magnitude below the current 
astrophysical upper limits. 

It follows from eqs. (\ref{theta}) and (\ref{B}) that if the resonance 
occurs not too close to the neutrinosphere ($r_{res}\aprge 60$ km) and  
$\mu B_0 \ll 10^{-2}$ eV (which, e.g., for $\mu = 10^{-12}\mu_B$ is 
fulfilled provided that $B_0 \le 10^{17}$ G), $\theta_i\approx \pi/2$. 
At the same time, we find that for
\EQ
\mu B_0 < 2.9 \times 10^{-6} \; {\rm eV}, \label{upper} 
\EN
the mixing angle close to the position of the stalled shock 
becomes very small, i.e. $\theta_f \approx 0$. The bound (\ref{upper}) 
was obtained assuming the resonance takes place at $r_{res}=350$ km; for 
smaller $r_{res}$ it gets relaxed. Comparing the upper bound (\ref{upper}) 
with the lower bound $\mu B_0 > 5.4 \times 10^{-8}$ eV which follows from 
eq. (\ref{lim2}) for $k=2$, we see that they do not contradict each other. 
If the condition (\ref{upper}) is not satisfied, the efficiency of the RSFP 
transition decreases, but for the adiabatic transitions ($\gamma \gg 1$) 
the transition probability never goes below 1/2. 
For $\theta_i\approx \pi/2$, $\theta_f\approx 0$ 
the transition probability is essentially determined by the degree of the 
adiabaticity of the transition.

In Fig. 2  we plot the transition probability $P$ vs $E/\Delta m^2$ for 
two different magnetic field configurations of eq. (\ref{B}), with $k=2$ 
and $k=3$. The parameter $\mu B_0$ was chosen in each case in such a way 
as to have the transition from the non-adiabatic to the adiabatic 
regime for the range of values of $E/\Delta m^2$ which corresponds to 
the resonance position between the neutrinosphere and the stalled shock. 
In what follows we shall assume that the adiabaticity condition 
(\ref{adiab}) and the condition (\ref{upper}) are satisfied, i.e. that the 
transition $\nu_{\mu(\tau)L} \rightarrow \bar\nu_{eR}$ is nearly complete. 

\section{Shock reheating}

After the bounce the energy of the shock is dissipated by the  
dissociation of nuclei, and the shock, after traveling up for some 
distance (which, following Fuller et al. \cite{Ful92}, we assume 
to be about $400$ km), stalls. Meanwhile neutrinos diffuse out from
the region where they were trapped; interacting with the matter 
behind the shock they deliver their energy in the neighborhood. The 
problem of shock reheating by neutrino interactions was discussed in 
detail by Bethe and Wilson \cite{BW85} (hereafter BW85). They considered 
the neutrino capture processes 
\EQ
\nu_e + n \rightarrow p + e^-, \;\;\; 
\bar\nu_e + p\rightarrow n + e^+ \label{react}
\EN
which the electron neutrinos and antineutrinos may undergo after diffusion 
from the supernova core; no neutrino flavor or spin--flavor transformation 
was taken into account. However the energy of the electron neutrinos 
is not sufficient to re-accelerate 
the shock. The possibility that neutrino flavor conversion due to the MSW 
effect may result in electron neutrinos having higher energies than that 
of the originally produced ones, thus increasing the amount of energy they 
deposit, was pointed out in \cite{MS86,Ful87} and then 
considered in detail in \cite{Ful92}. The authors of the latter paper came to 
the conclusion that the MSW effect in supernovae can give a net gain in the 
deposited energy of about 60\%. In this Section we estimate the gain in the 
neutrino--induced shock reheating energy due to the spin--flavor conversion of 
neutrinos. 

As discussed above, as a result of RSFP the 
$\mu$- and $\tau$- type neutrinos which are more energetic than
electron-type neutrinos will be converted into electron antineutrinos. 
These may then interact with the matter behind the shock more efficiently 
and deliver more energy than the originally produced $\bar{\nu}_e$'s. 
Here we estimate the energy gain due to the RSFP of supernova neutrinos 
following the consideration performed in \cite{Ful92} for the MSW effect. 

The neutrino absorption coefficients due to the reactions (\ref{react})  
can be written as 
\EQ
K_i (E_\nu)=N_A Y_i\langle \sigma (E_\nu)\rangle, \label{K1}
\EN 
where $N_A$ is Avogadro's number, $Y_i$ ($i=p,\,n$) is the appropriate 
nucleon number per baryon, and $\langle \sigma (E_\nu)\rangle$ is the 
reaction 
cross section averaged over the neutrino spectrum. Notice that the cross 
sections of reactions (\ref{react}) are essentially quadratic in neutrino 
energies 
and therefore $\langle \sigma (E_\nu)\rangle$ for the first and the second 
reactions is quadratic in the temperatures of electron neutrinos and 
antineutrinos, respectively. In what follows we will assume that the RSFP
is adiabatic for all the neutrino energies of interest ($\gamma \gg  1$), 
and so the conversion does 
not distort the neutrino spectra and just leads to their interchange. 
Assuming that the neutrino spectra are quasi-blackbody, the absorption 
coefficients can be approximated as \cite{BW85} 
\EQ
K_i(T_\nu)\approx (3.8 \times 10^{-19} \;{\rm cm}^2 \,{\rm g}^{-1})
\,Y_i T^2_\nu. \label{K2}
\EN
This takes into account the fact that there are actually fewer neutrinos 
with high energies than in the genuine blackbody spectrum. 

The rate at which the specific energy is deposited behind the shock is then 
\cite{BW85} 
\EQ
\dot E_{BW85}\approx (4\pi R_m)^{-2} [K_n(T_{\nu_e}) L_{\nu_e}+ 
K_p (T_{\bar\nu_e}) L_{\bar\nu_e}] - 4\pi j (T_m), \label{EBW}
\EN
where $L_{\nu_e}$ and $L_{\bar\nu_e}$ are the total $\nu_e$
and $\bar\nu_e$ luminosities, $R_m$ and $T_m$ the
radius and the temperature of the element of matter behind the 
shock, and $j(T_m)$ is the neutrino emissivity per steradian of the element
of matter considered. The last (negative) term in (\ref{EBW}) is negligible 
provided $T_m\ll T_{\nu_e}$. Following \cite{BW85} and \cite{Ful92}, in order 
to get a rough estimate of the effect we will first assume that this is the
case; we will come back to the discussion of the emissivity term later. 

The specific energy rate (\ref{EBW}) derived in \cite{BW85} 
does not take into account possible neutrino 
conversions. It is, however, straightforward to estimate
the gain in energy deposited by neutrinos behind the shock due to the 
RSFP transition $\nu_{\mu(\tau)L}\rightarrow \bar{\nu}_{eR}$. The ratio of 
the specific heating rates with and without RSFP transitions is 
\EQ
R_{RSFP}\equiv \frac{\dot E_{RSFP}}{\dot E_{BW85}}\approx
\frac{Y_n+Y_p\left(\frac {T_{\nu_\mu}}{T_{\nu_e}}\right)^2 }
     {Y_n+Y_p\left(\frac {T_{\bar\nu_e}}{T_{\nu_e}}\right)^2 }, \label{E} 
\EN
where we have taken into account that the total luminosities of all the 
neutrino species are approximately equal and that the cross sections of 
the two reactions in eq. (\ref{react}) for a given neutrino (antineutrino) 
energy $E_\nu$ are practically the same provided that $E_\nu \gg (m_n-m_p) 
\approx 1.3$ MeV. 

At the neutrino reheating epoch, $\langle E_{\nu_e}\rangle \approx 9$ 
MeV, $\langle E_{\bar\nu_e}\rangle \approx 12$ MeV and $\langle E_{\nu_\mu}
\rangle \approx\langle E_{\nu_\tau}\rangle \approx 20$ MeV \cite{Qian95}. 
Substituting these values into eq. (\ref{E}) and assuming that the reheating 
takes place at around 350 km where $Y_n\approx 0.53$, $Y_p\approx 0.47$ 
\cite{Ful92,Qian93}, one arrives at the following estimate:
\EQ
R_{RSFP} \approx 2.1. \label{Enum}
\EN

It is interesting to compare this with the analogous simple estimate 
of the energy gain due to the MSW effect. For the same values of $Y_p$, 
$Y_n$ and neutrino temperatures the gain factor would be $R_{MSW} \approx  
2.5$\footnote{Fuller et al. \cite{Ful92} obtained $R_{MSW}\approx 2$ for 
slightly different values of neutrino temperatures, $Y_p$ and $Y_n$. }, 
i.e. about 20\% larger than in the case of the RSFP transition. 
There are two reasons for this: first, the MSW effect would convert $\nu_\mu$ 
or $\nu_\tau$ into $\nu_e$ increasing the mean energy of the electron 
neutrinos by the factor $T_{\nu_\mu}/T_{\nu_e}\approx 2$, whereas the RSFP 
transition $\nu_{\mu(\tau)L}\to \bar{\nu}_{eR}$ increases the energy of 
electron antineutrinos by a smaller factor, $T_{\nu_\mu}/T_{\bar{\nu}_e} 
\approx 1.7$; second, the $\nu_e$'s which play a major role in the 
MSW--enhanced shock reheating interact with more abundant neutrons whereas 
the $\bar{\nu}_e$'s playing a major role in the RSFP--enhanced reheating 
interact with less abundant protons. Comparison of the two mechanisms shows 
that in general the RSFP--enhanced shock reheating is likely to be more 
efficient for smaller values of $\Delta m^2$ since the adiabaticity parameter 
(\ref{adiab}) becomes larger and the resonance occurs at lower densities, 
which means that the relative fraction of protons in the matter between the 
positions of the resonance and the stalled shock becomes larger. On the 
contrary, with increasing values of $\Delta m^2$ the resonance would occur at 
higher densities where the MSW transitions are more adiabatic and in addition 
the relative neutron abundance is higher, and so the MSW effect would become 
increasingly more important. 

It should be emphasized that our conclusion that the RSFP--induced neutrino 
conversions are typically slightly less efficient in reheating the shock 
than the MSW effect was the direct consequence of our assumption that 
$\Delta m^2>0$. However, neutrino masses do not have to follow the 
pattern of the charged fermion mass hierarchy; it is quite possible that 
they have an inverse hierarchy in which the $\nu_e$'s are the heaviest 
among the neutrinos (see, e.g., \cite{BerChk,PS,CM} for particle--physics 
models and \cite{RS,FPQ,PS,CM,R96} for discussions of possible 
phenomenological consequences). The neutrino mass hierarchy can be even more 
complicated 
with, e.g., the $\nu_e$ mass being in between $m_{\nu_\mu}$ and 
$m_{\nu_\tau}$ (in this case the MSW effect or the RSFP can be responsible 
for the solar neutrino deficit, while for $\nu_e$ being the heaviest neutrino 
they would not be operative in the sun). In any case, if $m_{\nu_e}$ is 
larger than at least one of the other neutrino masses, spin--flavor 
conversion of the type $\bar{\nu}_{\mu}(\bar{\nu}_{\tau})\to \nu_e$ will be 
resonantly enhanced in the supernova environment. This means that the 
resulting electron neutrinos will have a mean energy which is twice the 
energy of the originally produced $\nu_e$'s, and the gain in the shock 
reheating energy will be exactly the same as in the case of the MSW 
transitions. 
It should be noted, however, that because of the experimental upper limit 
$m_{\nu_e}<4.35$ eV (95\% c.l.) \cite{mnue}, the resonant $\bar{\nu}_{\mu}
(\bar{\nu}_{\tau})\to \nu_e$ conversion can only be relevant for supernova 
shock reheating for the electron neutrino mass lying in the relatively narrow 
range $\sim (3-4.3)$ eV. In addition, in this case the RSFP transition may be 
constrained severely by the supernova nucleosynthesis (r-process) 
arguments similar to those applied to the MSW effect in supernovae 
\cite{Qian93,Qian95,S}. At the same time, the RSFP--induced neutrino 
conversion with direct neutrino mass hierarchy may even help the supernova 
nucleosynthesis process \cite{FQN}. 

We would like to comment now on the implications of the RSFP--induced  
$\nu_\mu(\nu_\tau)\to \bar{\nu}_e$ conversion for the neutrino signals 
from supernovae. As we emphasized before, as a result of this conversion 
the $\bar{\nu}_e$'s emerging from the supernova would have higher energies 
than the originally produced ones. This should result in a ``stiffer'' 
than expected spectrum of electron antineutrinos observed through the 
reaction $\bar{\nu}_e+p\to n+e^+$ in the terrestrial water \v{C}erenkov 
detectors. Similar consequences would result for large-mixing-angle neutrino 
oscillations. Since the SN1987A $\bar{\nu}_e$  signals observed by Kamiokande 
and IMB detectors are in reasonable agreement with expectations, one might 
conclude that the conversions of $\nu_{\mu}(\nu_\tau)$ or their 
antiparticles into $\bar{\nu}_e$ are disfavored by the data \cite{SSB}. 
However, this conclusion relies heavily on the theoretical predictions for
the spectra of the supernova neutrinos and is therefore model dependent. 
In addition, though the SN1987A neutrino observations have confirmed the 
basic ideas of the supernova explosion and neutrino transport theory, 
the signals observed by the Kamiokande and IMB detectors are not fully 
understood. The opinions on whether a significant fraction of the SN1987A 
$\nu_{\mu}$'s or $\nu_\tau$'s or their antiparticles could have been 
converted into $\bar{\nu}_e$'s leading essentially to an interchange of 
their spectra differ significantly. The authors of \cite{SSB} and 
\cite{RS} believe that such a possibility is essentially ruled out, whereas 
the authors of \cite{FPQ,KK,QF95a} conclude that there is no useful limit on 
the probability of 
such an interchange. The authors of ref. \cite{JNR} turned the argument 
around and pointed out that if the solution of the solar neutrino problem 
through the MSW effect with the parameters that would lead to a 
significant 
interchange of the hard and soft neutrino spectra in the supernova is 
borne out by future solar neutrino experiments, one would have to 
conclude that the supernova $\nu_\mu(\nu_\tau)$ and/or $\bar{\nu}_e$ 
spectra are softer than had been thought previously. 

Without entering this discussion, we would like to point out a mechanism 
which could lead to the $\bar{\nu}_e$'s observed in the terrestrial 
detectors having the expected or even a softer spectrum even if a strong 
RSFP--induced $\nu_\mu (\nu_\tau)\to \bar{\nu}_{e}$ transition occurs in 
the supernova. The possibility of having a softer than expected 
$\bar{\nu}_e$ spectrum is especially interesting since it is favored by 
the SN1987A data. It has been shown in \cite{APS1} that, if the neutrinos 
have non-zero vacuum mixing angle in addition to the transition magnetic 
moment, and the direction of the transverse magnetic field changes along 
the neutrino path, resonant $\bar{\nu}_e \leftrightarrow \nu_e$ 
transitions are possible. In this case the $\bar{\nu}_e$'s observed in 
the terrestrial detectors may have the spectra of the originally 
produced $\nu_e$'s or $\bar{\nu}_e$'s, depending on the history of 
the conversions that the neutrinos experienced in the supernova before 
they reach the $\bar{\nu}_e \leftrightarrow \nu_e$ resonance. This 
possibility will be discussed in more detail elsewhere. 

Our estimate of the energy gain due to the RSFP of supernova 
neutrinos was in fact very rough. More accurate calculations would 
involve integration of the energy deposition rate over the entire region  
between the resonance and the position of the shock, and should take into 
account neutrino re-emission by matter [the last term in eq. (\ref{EBW})]. 
Such a calculation was carried out in \cite{Ful92} for the MSW--enhanced  
shock reheating. The result was about a factor 0.8 decrease of the energy 
gain ratio $R_{MSW}$ as compared with the simple estimate based on the 
formula similar to our eq. (\ref{E}). The main reason for this decrease is 
that MSW--enhanced shock reheating increases the local temperature of 
the matter $T_m$ and thus the emissivity term in eq. (\ref{EBW}), 
resulting in a 
``negative feedback'' for the effect \cite{Ful92}. We expect a similar 
reduction to take place for the RSFP--enhanced reheating, and therefore the 
estimate of eq. (\ref{E}) should be replaced by $R_{RSFP}\simeq$ 
(1.6--1.7).  

\section{Conclusions}
We have shown that resonant spin--flavor precession of neutrinos due to 
interaction of their transition magnetic moments with the strong 
magnetic fields inside supernovae may increase the energy deposited 
by neutrinos in the matter behind the shock by about 60\% and thus 
help to re-accelerate it. 
For the process to be efficient in the supernova environment, the heavier  
neutrino mass should be in the range $\sim (3 - 600)$ eV, and the transition 
magnetic moment $\mu$ should be of the order of $10^{-14}\mu_B$ provided 
that the magnetic field strength at the resonance position is of the order of 
$10^{12}-10^{15}$ G. All these values of the parameters are consistent with 
the available laboratory and astrophysical constraints on neutrino 
properties as well as our present ideas about supernova magnetic fields. 
In fact, the neutrinos with the masses and magnetic moments in the above 
range may be very interesting for cosmology and for the decaying neutrino 
theory \cite{Sciama1993book}.  

Our simple consideration gave only a rough estimate of the supernova 
explosion energy increase due to the RSFP conversion; whether or not 
this effect is sufficient to revive the shock leading to a successful 
supernova explosion can only be decided on the basis of a full-scale 
supernova dynamics calculation with the RSFP transition included, which goes 
beyond the scope of the present paper. However, our estimates show that 
the RSFP--induced neutrino conversion can result in quite a sizable 
increase of the supernova explosion energy, and we believe that this 
effect deserves further investigation. 

E.A. is grateful to Alexei Smirnov for useful discussions and to 
George Fuller and Hiroshi Nunokawa for correspondence. S.T.P. would 
like to thank George Fuller and Yong Qian for discussions of the 
physics of supernovae. E.A. is grateful to ICTP and SISSA, where part of this 
work has been done, for hospitality and support. The work of A.L. and D.W.S. 
has been supported by MURST.

\newpage
\centerline{\bf \large Figure captions}
\vskip 1truecm
\noindent
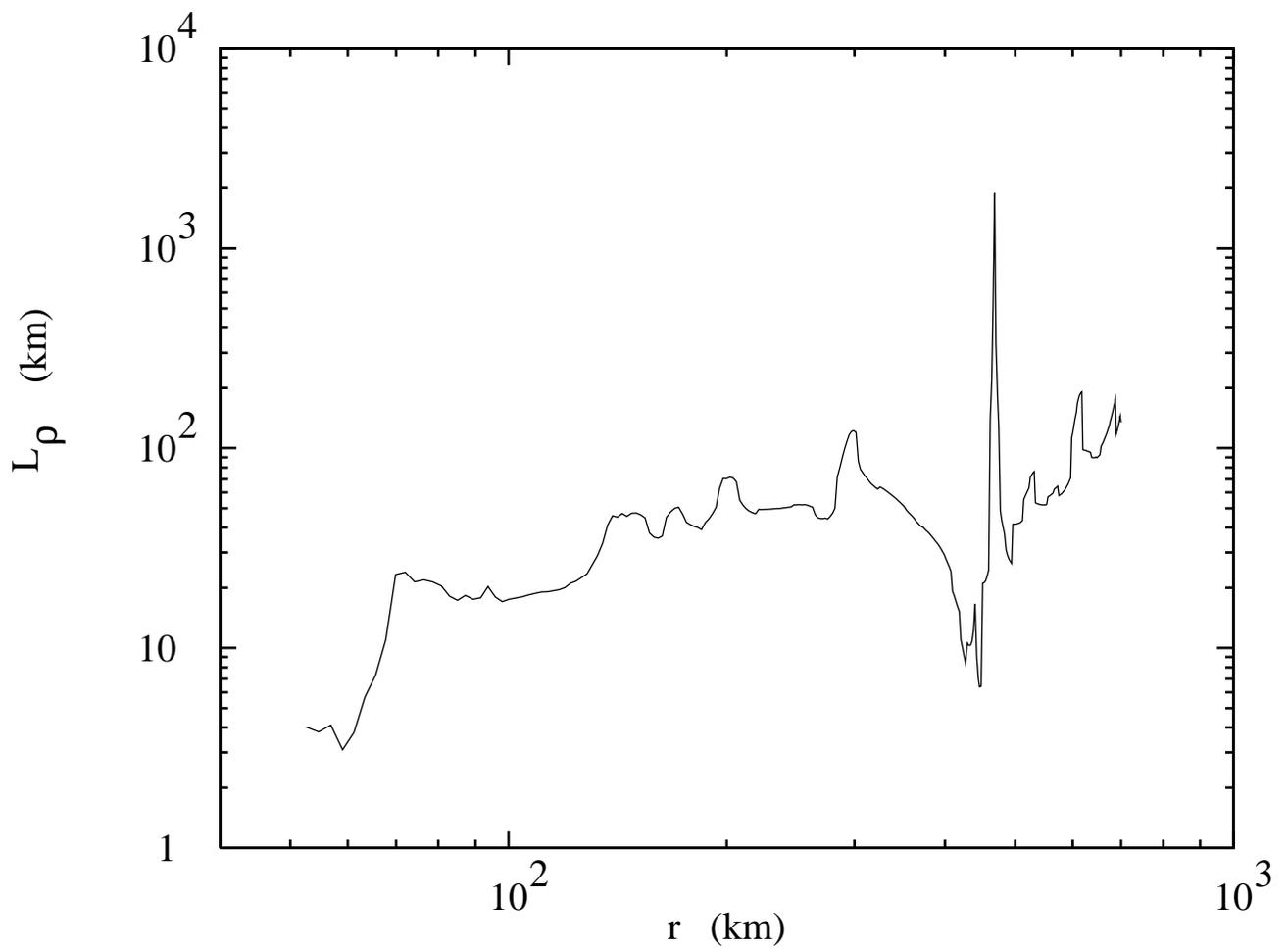
Fig. 1. The characteristic scale height $L_\rho$ as a function of the 
radial distance $r$ from the center of the supernova at $t\approx 0.15$ s 
after the bounce. The shock wave is located at $r \approx 430$ km. The matter 
density and $Y_e$ profiles of refs. \cite{Ful92,Qian93} have been used.

\vskip 0.5truecm
\noindent
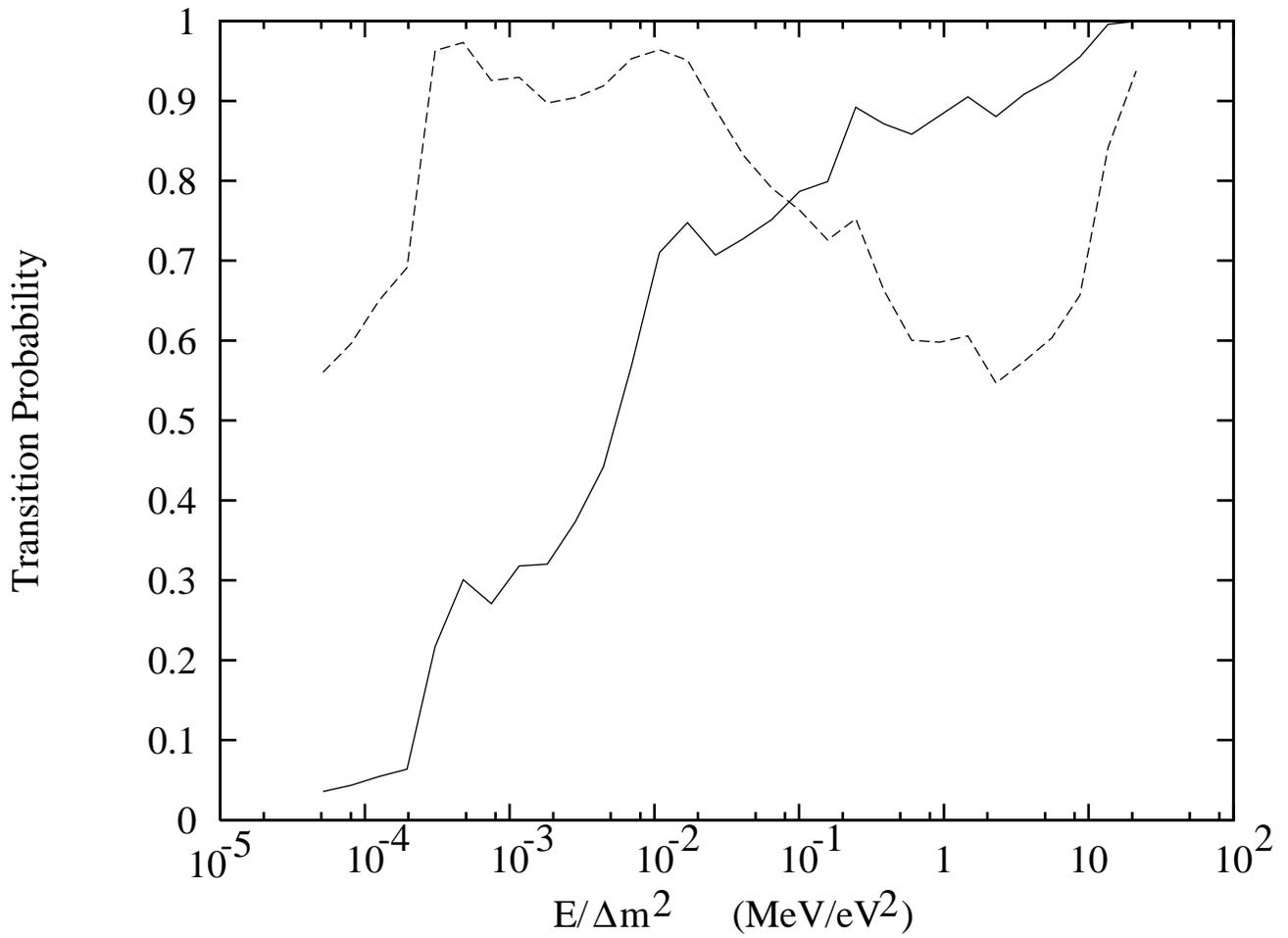
Fig. 2. The transition probability $P$ versus $ E/\Delta m^2$ for two 
different magnetic field distributions. The solid line corresponds to a 
power law [cf. eq. (\ref{B})] with $k=2$ and $\mu B_0=5.8\times 10^{-8}$, 
eV, the dashed line, to a power law with $k=3$ and $\mu B_0=2.9 \times 
10^{-7}$ eV. 

\end{document}